\documentclass[10pt, conference]{IEEEtran}

\makeatletter
\def\ps@headings{%
\def\@oddhead{\mbox{}\scriptsize\rightmark \hfil \thepage}%
\def\@evenhead{\scriptsize\thepage \hfil \leftmark\mbox{}}%
\def\@oddfoot{}
\def\@evenfoot{}}
\def\blfootnote{\xdef\@thefnmark{}\@footnotetext}
\makeatother
\usepackage{graphicx} 
\usepackage{epsfig} 
\usepackage{color}
\usepackage{mathptmx} 
\usepackage{times} 
\usepackage{latexsym}
\usepackage{amsmath} 
\usepackage{amssymb}  
\usepackage{subfigure}
\usepackage{algorithmic}
\usepackage{setspace}
\usepackage[bottom]{footmisc}
\newtheorem{theorem}{Theorem}
\newtheorem{conjecture}{Conjecture}
\newtheorem{lemma}{Lemma}
\newtheorem{definition}{Definition}

\title{\huge \bf
Optimal ordering of transmissions for computing Boolean threhold functions
}                                                                     

\author{\IEEEauthorblockN{Hemant Kowshik}
\IEEEauthorblockA{CSL and Department of ECE\\
University of Illinois Urbana-Champaign\\
Email: kowshik2@illinois.edu}
\and
\IEEEauthorblockN{P. R. Kumar}
\IEEEauthorblockA{CSL and Department of ECE\\
University of Illinois Urbana-Champaign\\
Email: prkumar@illinois.edu}

}
\begin{document}
\maketitle\blfootnote{This material is based upon work partially supported by AFOSR under Contract FA9550-09-0121, NSF under Contract No. CNS-07-21992, and USARO under Contract Nos. W911NF-08-1-0238 and W-911-NF-0710287.
Any opinions, findings, and conclusions or recommendations expressed in this publication are those of the authors and do not necessarily reflect the views of the above agencies.}
\thispagestyle{empty}
\pagestyle{empty}
\begin{abstract}
We address a sequential decision problem that arises in the computation of symmetric Boolean functions of distributed data. We consider a collocated network, where each node's transmissions can be heard by every other node. Each node has a Boolean measurement and we wish to compute a given Boolean function of these measurements. We suppose that the measurements are independent and Bernoulli distributed. Thus, the problem of optimal computation becomes the problem of optimally ordering node's transmissions so as to minimize the total expected number of bits. 

We solve the ordering problem for the class of Boolean threshold functions. The optimal ordering is dynamic, i.e., it could potentially depend on the values of previously transmitted bits. Further, it depends only on the ordering of the marginal probabilites, but not on their exact values. This provides an elegant structure for the optimal strategy. For the case where each node has a block of measurements, the problem is significantly harder, and we conjecture the optimal strategy. 
\end{abstract}
\section{INTRODUCTION}
Most sensor network applications are typically interested  only in computing some relevant \textit{function} of the correlated data at distributed sensors. For example, one might want to compute the mean temperature for environmental monitoring, or the maximum temperature in fire alarm systems. On the other hand, sensor nodes are severely limited in terms of power and bandwidth, and are generating enormous quantities of data. Thus, we seek efficient in-network computation and communication strategies for the function of interest. 


Computing and communicating functions of distributed data presents several challenges. On the one hand, the wireless medium being a broadcast medium, nodes have to deal with interference from other transmissions. On the other hand, nodes can exploit these overheard transmissions, and the structure of the function to be computed, to achieve a more efficient description of their own data.  Moreover, the strategy for computation may benefit from interactive information exchange between nodes. 

We consider a collocated network where each node's transmissions can be heard by every other node. 
At most one node is allowed to transmit successfully at any time. Each node has a Boolean variable and we focus on the specific problem of symmetric Boolean function computation. We adopt a deterministic formulation of the problem of function computation, allowing zero error. We suppose that node measurements are independent and distributed according to given marginal Bernoulli distributions. In this paper, we focus on optimal strategies for Boolean threshold functions, which are equal to $1$ if and only if the number of nodes with measurement $1$ is greater than a certain threshold. The set of admissible strategies includes all interactive strategies, where a node may exchange several messages with other nodes.

In the case where each node has  a single bit, the communication problem is rendered trivial, since it is optimal for the transmitting node to simply indicate its bit value. Thus, it only remains to determine the optimal ordering of nodes' transmissions so as to minimize the expected number of bits exchanged. For the class of Boolean threshold functions, we present a simple indexing policy for ordering the transmissions and prove its optimality. The optimal policy is dynamic, possibly depending on the previously transmitted bits. Further, the optimal policy depends only on the ordering of the marginal probabilities, but surprisingly not on their values.

The problem of optimally ordering transmissions of nodes is a sequential decision problem and can indeed be solved by dynamic programming. However, this would require solving the dynamic program for all thresholds and all probability distributions, which is computationally hard. We avoid this, and establish a more insightful solution, in the form of a simple rule defining the optimal policy.

In Section \ref{sec_one}, we formulate the problem of single instance computation, and derive the resulting dynamic programming equation. We then propose the indexing policy and present a detailed proof of optimality, by induction on the number of nodes in the network. 
In Section \ref{sec_two}, we consider the extension to the case of block computation, where each node has a block of measurements and we are allowed block coding. This problem is significantly harder, 
and we conjecture the structure of an optimal multi-round policy, building on the optimal policy for single instance computation. 
\section{RELATED WORK}
The the problem of worst-case block function computation with zero error was formulated  in \cite{GiridharKumar}. 
The authors identify two classes of symmetric functions namely \textit{type-sensitive} functions exemplified by Mean, Median and Mode, and \textit{type-threshold} functions, exemplified by Maximum and Minimum. The maximum rates for computation of type-sensitive and type-threshold functions in random planar networks are shown to be $\Theta(\frac{1}{\log n})$ and $\Theta(\frac{1}{\log \log n})$ respectively, where $n$ is the number of nodes. If we impose a probability distribution on the node measurements, one can show that the average case complexity of computing type-threshold functions is $\Theta(1)$ \cite{KowshikKumar}.

In this paper, we require that every node must compute the function. This approach naturally allows the use of tools from communication complexity \cite{KushiNisan}. In communication complexity \cite{KushiNisan}, we seek to find the minimum number of bits that must be exchanged between two nodes to achieve worst-case zero-error computation of a function of the node variables. The problem of worst-case Boolean function computation was first considered in \cite{AhlswedeCai}, where the complexity of the Boolean AND function was shown to be $\log_{2}3$ bits. In \cite{KowshikKumar_ITW}, this was considerably generalized to derive the exact complexity of computing Boolean threshold functions. 

If the measurements are drawn from some joint probability distribution and one is allowed block computation, we arrive at a distributed source coding problem with a fidelity criterion that is function-dependent, for which little is known. One special case, a source coding problem for function computation with side information, has been studied in \cite{OrlitskyRoche}. The problem of interactive function computation in collocated networks has been studied in \cite{MaGuptaIshwar}.

The problem of minimizing the depth of decision trees for Boolean threshold queries is considered in \cite{AsherNewman}. In \cite{ArrowPesotSobel}, an interesting problem in sequential decision making is studied, where, $n$ nodes have i.i.d. measurements, and a central agent wishes to know the identities of the nodes with the $k$ largest values. One is allowed questions of the type ``Is $X \geq t$'', to which the central agent receives the list of all nodes which satisfy the condition. Under this framework, the optimal recursive strategy of querying the nodes is found. A key difference in our formulation is that we are only allowed to query particular nodes, and not all nodes at once.
\section{Optimal ordering for single instance computation}\label{sec_one}
Consider a collocated network with nodes $1$ through $n$, where each node $i$ has a Boolean measurement $X_i \in \{0,1\}$. The $X_i$s are independent of each other and drawn from a Bernoulli distribution with $P(X_i = 1) = : p_i$. Without loss of generality, we assume that $p_1 \leq p_2 \leq \cdots \leq p_n$.  

We address the following problem. \textit{Every} node wants to compute the same function $f(X_1, X_2, \ldots, X_n)$ of the measurements. We seek to find communication schemes which achieve correct function computation at each node, with minimum expected total number of bits exchanged. Throughout this paper, we consider the broadcast scenario where each node's transmission can be heard by every other node. We also suppose that collisions do not convey information thus restricting ourselves to \textit{collision-free strategies} as in \cite{GiridharKumar}. This means that for the $k^{th}$ bit $b_k$, the identity of the transmitting node $T_k$ depends only on previously broadcast bits $b_1, b_2, \ldots, b_{k-1}$, while the value of the bit it sends can depend arbitrarily on all previous broadcast bits as well as its own measurements $X_{T_k}$.

First, we note that since each node has exactly one bit of information, it is optimal to set $b_k = X_{T_k}$. Indeed, for any other choice $b_k' = g(b_1, \ldots, b_{k-1}, X_{T_{k}})$, the remaining nodes can reconstruct $b_k'$ since they already know $b_i, \ldots, b_{k-1}$. Thus the only freedom available is in choosing the transmitting node $T_k$ as a function of $b_1, b_2, \ldots, b_{k-1}$, for otherwise the transmission itself could be avoided. We call this the \textit{ordering problem}. Thus, by definition, the order can dynamically depend on the previous broadcast bits. In this paper, we address the ordering problem for a class of Boolean functions, namely threshold functions.

\textbf{Notation:} The set of measurements of nodes $1$ through $n$ is denoted by $(X_1, X_2, \ldots, X_n)$ which is abbreviated as $\mathbf{X}^{n}$. In the sequel, we will use $\mathbf{X}^{n}_{-i}$ to denote the set of measurements $(X_1, \ldots, X_{i-1}, X_{i+1}, \ldots, X_n)$. As a natural extension, we use $\mathbf{X}^{n}_{-( i,j)}$ to denote the set of measurements $(X_1, \ldots, X_{i-1}, X_{i+1}, \ldots, X_{j-1}, X_{j+1}, \ldots, X_n)$, where $i < j$.

\subsection{Optimal ordering for computing Boolean threshold functions}
\begin{definition}[Boolean threshold functions]
A Boolean threshold function $\Pi_{\theta}(X_1, X_2, \ldots, X_n)$ is defined as
\begin{displaymath}
\Pi_{\theta}(X_1, X_2, \ldots, X_n) = \left\{ \begin{array}{l} 1 \quad \textrm{if } \sum_{i}X_i \geq \theta , \\ 0 \quad \textrm{otherwise.}\end{array}\right.
\end{displaymath}
\end{definition}

Given a function $\Pi_{n-k}(\mathbf{X}^{n})$, the ordering problem can indeed be solved using dynamic programming. Let $C(\Pi_{n-k}(\mathbf{X}^{n})$ denote the minimum expected number of bits required to compute $\Pi_{n-k}(\mathbf{X}^{n})$. The dynamic programming equation is

\vspace{-0.1in}\begin{small}
\begin{displaymath} \label{dyn_prog_eqn}
C(\Pi_{n-k}(\mathbf{X}^{n})) = \min_{i}\{1 + p_i C(\Pi_{n-k-1}(\mathbf{X}^{n}_{-i})) + (1-p_i)C(\Pi_{n-k}(\mathbf{X}^{n}_{-i}))\}.
\end{displaymath}
\end{small}
However solving this equation is computationally complex. Further, it is unclear at the outset if the optimal strategy will depend only on the ordering of the $p_i$s, or their particular values. This makes the explicit solution of (\ref{dyn_prog_eqn}) for all $n$, $k$ and $(p_1, p_2, \ldots p_n)$ notoriously hard. We present a very simple characterization of the optimal strategy for each $n$ and $0 \leq k \leq n-1$ and show that this is independent of the particular values of the $p_i$s, but only depends on the ordering. 

To begin with, we argue that solving the ordering problem for Boolean threshold functions, is equivalent to solving the following problem for each $n$ and $k$: In the optimal strategy for computing $\Pi_{n-k}(X_1, X_2, \ldots X_n)$ determine which node must transmit first. Indeed, if $T(1)$ is the first node to transmit under the optimal strategy, then, depending on whether $X_{T(1)} = 0$ or $X_{T(1)} = 1$, the rest of the nodes would need to compute $\Pi_{n-k}(\mathbf{X}^{n}_{-T(1)})$ or $\Pi_{n-k-1}(\mathbf{X}^{n}_{-T(1)})$. Since we solved the problem for all $n$ and $k$, we can determine which node should transmit next in either case. 
\begin{theorem}\label{thm_seq_bool_threshold}
In order to compute the Boolean threshold function $\Pi_{n-k}(\mathbf{X}^{n})$, it is optimal for node $k+1$ to transmit first. This result is true for all $n$ and all $0 \leq k \leq n-1$ and all probability distributions with $p_1 \leq p_2 \leq \ldots \leq p_n$.
\end{theorem}
\textbf{Proof:} Define $\tilde{C}(\Pi_{n-k}(\mathbf{X}^{n})) := C(\Pi_{n-k}(\mathbf{X}^{n})) - 1$ for notational convenience. We also define the following expressions.

\vspace{-0.1in}\begin{footnotesize}
\begin{multline}
\hspace{-0.1in}T_{m, k, i}(\mathbf{X}^{m}) := p_{k+1}\tilde{C}(\Pi_{m-k-1}(\mathbf{X}^{m}_{-(k+1)})) + (1-p_{k+1})\tilde{C}(\Pi_{m-k}(\mathbf{X}^{m}_{-(k+1)})) \\
\qquad - p_{i}\tilde{C}(\Pi_{m-k-1}(\mathbf{X}^{m}_{-i})) - (1-p_{i})\tilde{C}(\Pi_{m-k}(\mathbf{X}^{m}_{-i})). \label{T_defn}
\end{multline}
\end{footnotesize}
$T_{m, k, i}$ is the difference between the expected number of bits when node $k+1$ transmits first, and the expected number of bits when node $i$ transmits first.

\vspace{-0.1in}\begin{small}
\begin{multline}
\hspace{-0.1in}S^{(1)}_{m,k,i}(\mathbf{X}^{m}) := (p_{k+1} - p_{i})C(\Pi_{m-k-1}(\mathbf{X}^{m}_{-(k+1, i)})) \\
+ (1-p_{k+1})\tilde{C}(\Pi_{m-k}(\mathbf{X}^{m}_{-(k+1)})) - (1-p_{i})\tilde{C}(\Pi_{m-k}(\mathbf{X}^{m}_{-i})) \label{S1_defn}
\end{multline}
\begin{multline} 
S^{(2)}_{m,k,i}(\mathbf{X}^{m}) := (p_{i} - p_{k+1})C(\Pi_{m-k-1}(\mathbf{X}^{m}_{-(i, k+1)})) \\
+ p_{k+1}\tilde{C}(\Pi_{m-k-1}(\mathbf{X}^{m}_{-(k+1)})) - p_{i}\tilde{C}(\Pi_{m-k-1}(\mathbf{X}^{m}_{-i})) \label{S2_defn}
\end{multline}
\end{small}
We do not yet have an interpretation for $S^{(1)}_{m,k,i}$ and $S^{(2)}_{m,k,i}$. However, we will use these expressions in the sequel.

We establish the above theorem by induction on the number of nodes $n$. However, we need to load the induction hypothesis. Consider the following induction hypothesis.
\begin{eqnarray}
\textrm{(a) } T_{m,k,i}(\mathbf{X}^{m}) & \leq & 0 \quad \textrm{for all } 0 \leq k \leq (m-1), 1 \leq i \leq m  \nonumber\\
\textrm{(b) } S^{(1)}_{m,k,i}(\mathbf{X}^{m}) & \leq & 0 \quad \textrm{for all } 0 \leq k \leq (m-1), k+2 \leq i \leq m  \nonumber \\
\textrm{(c) } S^{(2)}_{m,k,i}(\mathbf{X}^{m}) & \leq & 0 \quad \textrm{for all } 0 \leq k \leq (m-1), 1 \leq i \leq k  \nonumber
\end{eqnarray}
Observe that part $(a)$ immediately establishes that $k+1$ should transmit first in the optimal strategy for computing the function $\Pi_{m-k}(\mathbf{X}^{m})$.

The basis step for $m = 1, k = 1$ is trivially true. Let us suppose the induction hypothesis is true for all $m \leq n$. We now proceed to prove the hypothesis for $m = n+1$. 
\begin{lemma}\label{lemma_three}
For fixed $k$ and $i \geq k+2$, we have $S^{(1)}_{n+1,k,i}(\mathbf{X}^{n+1}) \leq 0$.
\end{lemma}
\textbf{Proof:} See Appendix \ref{sec_proof_three}.
\begin{lemma}\label{lemma_four} 
For fixed $k$ and $i \leq k$, we have $S^{(2)}_{n+1,k,i}(\mathbf{X}^{n+1}) \leq 0$.
\end{lemma}
\textbf{Proof:} See Appendix \ref{sec_proof_four}.

Lemmas \ref{lemma_three} and \ref{lemma_four} establish the induction step for parts $(b)$ and $(c)$ of the induction hypothesis. We now proceed to show the induction step for part $(a)$.
\begin{lemma}\label{lemma_one}
For fixed $k$ and $i \geq k+2$, we have $T_{n+1, k, i}(\mathbf{X}^{n+1}) \leq S^{(1)}_{n+1,k,i}(\mathbf{X}^{n+1})$.
\end{lemma}
\textbf{Proof:} See Appendix \ref{sec_proof_one}
\begin{lemma}\label{lemma_two}
For fixed $k$ and $i \leq k$, we have $T_{n+1, k, i}(\mathbf{X}^{n+1}) \leq S^{(2)}_{n+1, k, i}(\mathbf{X}^{n+1})$.
\end{lemma}
\textbf{Proof:} See Appendix \ref{sec_proof_two}

Using Lemmas \ref{lemma_one} and \ref{lemma_two} together with Lemmas \ref{lemma_three} and \ref{lemma_four}, we see that $T_{n+1, k, i}(\mathbf{X}^{n+1}) \leq 0$ for all $0 \leq k \leq n$ and $i \neq k+1$. For the case $i = k+1$, we have $T(n+1, k, k+1) = 0$ trivially. This completes the induction step for part $(a)$, and the proof of the Theorem. $\Box$
\section{Optimal ordering for block computation}\label{sec_two}
We now shift attention to the case where we allow for nodes to accumulate a block of $N$ measurements, and thus achieve improved efficiency by using block codes. We consider the class of all interactive strategies for computation, where the $k$th bit can depend arbitrarily on all previously broadcast bits. We require that all nodes compute the function with zero error for the block. We present a conjecture for the optimal strategy based on the insight gained from the single instance solution. 
\begin{conjecture}
In order to compute the Boolean threshold function $\Pi_{n-k}(\mathbf{X}^{n})$, it is optimal for node $k+1$ to transmit first, using the Huffman code. This result is true for all $n$ and all $0 \leq k \leq n-1$ and all probability distributions with $p_1 \leq p_2 \leq \ldots \leq p_n$.
\end{conjecture}

Observe that after node $k+1$ transmits, we are left with two block computation problems. For the instances where $X_{k+1} = 0$, we need to compute $\Pi_{n-k}(\mathbf{X}^{n}_{-(k+1)})$ and for the instances where $X_{k+1} = 1$, we need to compute $\Pi_{n-k-1}(\mathbf{X}^{n}_{-(k+1)})$. Thus the conjectured strategy can be recursively applied, yielding an interactive multi-round strategy. However, proving the optimality of this strategy is significantly harder. For worst case block computation, the lower bound is established using fooling sets \cite{KowshikKumar_ITW}. Adapting this idea to the probabilistic scenario remains an interesting challenge for the future. 

\section{Concluding remarks}
We have considered a sequential decision problem, that arises in the context of optimal computation of Boolean threshold functions in collocated networks. For single instance computation, we show that the optimal strategy has an elegant structure, which depends only on the ordering of the marginal probabilities, and not on their exact values. The extension to the case of block computation is harder and remains a challenge for the future.  It is also interesting to extend this result to the case of correlated measurements
\begin{figure*}
\appendix
\section{Proofs}
\subsection{Proof of Lemma \ref{lemma_three}}\label{sec_proof_three}
\begin{eqnarray}
\hspace{-0.2in}& & (p_{k+1} - p_{i})C(\Pi_{n-k}(\mathbf{X}^{n+1}_{-(k+1, i)})) + (1-p_{k+1})\tilde{C}(\Pi_{n-k+1}(\mathbf{X}^{n+1}_{-(k+1)})) - (1-p_{i})\tilde{C}(\Pi_{n-k+1}(\mathbf{X}^{n+1}_{-i})) \nonumber \\
\hspace{-0.2in}& = & (p_{k+1} - p_i)\left[ 1 + p_k C(\Pi_{n-k-1}(\mathbf{X}_{-(k, k+1, i)}^{n+1})) + (1-p_k) C(\Pi_{n-k}(\mathbf{X}^{n+1}_{-(k, k+1, i)}))\right]  \nonumber \\
\hspace{-0.2in}& &  + (1-p_{k+1})\left[p_k C(\Pi_{n-k}(\mathbf{X}^{n+1}_{-(k, k+1)})) + (1- p_k)C(\Pi_{n-k+1}(\mathbf{X}^{n+1}_{-(k, k+1)}))\right] \nonumber \\
\hspace{-0.2in}& & -(1- p_i)\left[p_k C(\Pi_{n-k}(\mathbf{X}^{n+1}_{-(k, i)})) + (1- p_k) C(\Pi_{n-k+1}(\mathbf{X}^{n+1}_{-(k, i)}))\right] \label{proof_three_a} \\
\hspace{-0.2in}& = & p_k \left[(p_{k+1} - p_i)C(\Pi_{n-k-1}(\mathbf{X}_{-(k, k+1, i)}^{n+1})) \right.  + \left. (1-p_{k+1}) \tilde{C}(\Pi_{n-k}(\mathbf{X}^{n+1}_{-(k, k+1)})) - (1-p_i)\tilde{C}(\Pi_{n-k}(\mathbf{X}^{n+1}_{-(k, i)}))\right] \nonumber \\
\hspace{-0.2in}& & +(1-p_k)\left[(p_{k+1} - p_i)C(\Pi_{n-k}(\mathbf{X}_{-(k, k+1, i)}^{n+1})) \right. + \left. (1-p_{k+1}) \tilde{C}(\Pi_{n-k+1}(\mathbf{X}^{n+1}_{-(k, k+1)})) - (1-p_i)\tilde{C}(\Pi_{n-k+1}(\mathbf{X}^{n+1}_{-(k, i)}))\right]  \nonumber \\
\hspace{-0.2in}& \leq & p_k \left[(p_{k+1} - p_i)C(\Pi_{n-k-1}(\mathbf{X}_{-(k, k+1, i)}^{n+1})) + (1-p_{k+1}) \tilde{C}(\Pi_{n-k}(\mathbf{X}^{n+1}_{-(k, k+1)}))\right. - \left.(1-p_i)\tilde{C}(\Pi_{n-k}(\mathbf{X}^{n+1}_{-(k, i)}))\right] + (1-p_k)S^{(1)}_{n, k-1, i-1}(\mathbf{X}^{n+1}_{-k}) \nonumber \\
\hspace{-0.2in}& \leq & p_k \left[(p_{k+1} - p_i)C(\Pi_{n-k-1}(\mathbf{X}_{-(k, k+1, i)}^{n+1})) + (1-p_{k+1}) \tilde{C}(\Pi_{n-k}(\mathbf{X}^{n+1}_{-(k, k+1)}))\right. - \left. (1-p_i)\tilde{C}(\Pi_{n-k}(\mathbf{X}^{n+1}_{-(k, i)}))\right] \label{proof_three_b} \\
\hspace{-0.2in}& = & p_k \left[(p_{k+1} - p_i) C(\Pi_{n-k-1}(\mathbf{X}^{n+1}_{-(k, k+1, i)})) + (1-p_{k+1}) \tilde{C}(\Pi_{n-k}(\mathbf{X}^{n+1}_{-(k, k+1)}))\right.  \nonumber \\
\hspace{-0.2in} & & \left. -(1-p_i)[p_{k+1} C(\Pi_{n-k-1}(\mathbf{X}^{n+1}_{-(k,k+1, i)})) + (1-p_{k+1}) C(\Pi_{n-k}(\mathbf{X}^{n+1}_{-(k, k+1, i)}))]\right] \label{proof_three_c} \\
\hspace{-0.2in}& = & p_k(1-p_{k+1})\left[\tilde{C}(\Pi_{n-k}(\mathbf{X}^{n+1}_{-(k, k+1)})) - p_i C(\Pi_{n-k-1}(\mathbf{X}^{n+1}_{-(k, k+1, i)}))\right.  - \left.(1-p_i) C(\Pi_{n-k}(\mathbf{X}^{n+1}_{-(k, k+1, i)}))\right] \leq 0. \label{proof_three_d} 
\end{eqnarray}
Equation (\ref{proof_three_a}) follows from the optimal ordering for computing $\Pi_{n-k}(\mathbf{X}^{n+1}_{-(k+1,i)})$, $\Pi_{n-k+1}(\mathbf{X}^{n+1}_{-(k+1)})$ and $\Pi_{n-k+1}(\mathbf{X}^{n+1}_{-i})$, which is true by the induction hypothesis for $m = n$. The inequality (\ref{proof_three_b}) follows from the induction hypothesis that $S^{(1)}_{n,k-1,i}(\mathbf{X}^{n+1}_{-k}) \leq 0$. Equality in (\ref{proof_three_c}) and (\ref{proof_three_d}) follows from the optimal ordering for computing $\Pi_{n-k}(\mathbf{X}^{n+1}_{-(k, i)})$  and $\Pi_{n-k}(\mathbf{X}^{n+1}_{-(k, k+1)})$ respectively. $\Box$
\end{figure*}
\begin{figure*}
\subsection{Proof of Lemma \ref{lemma_four}}\label{sec_proof_four}
\begin{eqnarray}
\hspace{-0.2in}& & (p_{i} - p_{k+1})C(\Pi_{n-k}(\mathbf{X}^{n+1}_{-(i, k+1)})) + p_{k+1}\tilde{C}(\Pi_{n-k}(\mathbf{X}^{n+1}_{-(k+1)})) - p_i\tilde{C}(\Pi_{n-k}(\mathbf{X}^{n+1}_{-i})) \nonumber \\
\hspace{-0.2in}& = & (p_{i} - p_{k+1})\left[ 1 + p_{k+2} C(\Pi_{n-k-1}(\mathbf{X}_{-(i, k+1, k+2)}^{n+1})) \right. \nonumber + \left. (1-p_{k+2}) C(\Pi_{n-k}(\mathbf{X}^{n+1}_{-(i, k+1, k+2)}))\right] \nonumber \\
\hspace{-0.2in}& &+ p_{k+1}\left[p_{k+2} C(\Pi_{n-k-1}(\mathbf{X}^{n+1}_{-(k+1, k+2)})) + (1- p_{k+2})C(\Pi_{n-k}(\mathbf{X}^{n+1}_{-(k+1, k+2)}))\right] \nonumber \\
\hspace{-0.2in}& &- p_i\left[p_{k+2} C(\Pi_{n-k-1}(\mathbf{X}^{n+1}_{-(i, k+2)})) + (1- p_{k+2}) C(\Pi_{n-k}(\mathbf{X}^{n+1}_{-(i, k+2)}))\right] \label{proof_four_a} \\
\hspace{-0.2in}& = & p_{k+2} \left[(p_{i} - p_{k+1})C(\Pi_{n-k-1}(\mathbf{X}_{-(i, k+1, k+2)}^{n+1})) \right. + p_{k+1} \tilde{C}(\Pi_{n-k-1}(\mathbf{X}^{n+1}_{-(k+1, k+2)})) - \left. p_i\tilde{C}(\Pi_{n-k-1}(\mathbf{X}^{n+1}_{-(i, k+2)}))\right] \nonumber \\
\hspace{-0.2in}& & +(1-p_{k+2})\left[(p_{i} - p_{k+1})C(\Pi_{n-k}(\mathbf{X}_{-(i, k+1, k+2)}^{n+1})) \right. + p_{k+1} \tilde{C}(\Pi_{n-k}(\mathbf{X}^{n+1}_{-(k+1, k+2)})) - \left. p_i \tilde{C}(\Pi_{n-k}(\mathbf{X}^{n+1}_{-(i, k+2)}))\right] \nonumber \\
\hspace{-0.2in}& \leq &  p_{k+2}\left[S^{(2)}_{n, k, i}(\mathbf{X}^{n+1}_{-(k+2)})\right] + (1-p_{k+2})\left[(p_{i} - p_{k+1})C(\Pi_{n-k}(\mathbf{X}_{-(i, k+1, k+2)}^{n+1}))\right. + \left. p_{k+1} \tilde{C}(\Pi_{n-k}(\mathbf{X}^{n+1}_{-(k+1, k+2)})) - p_i \tilde{C}(\Pi_{n-k}(\mathbf{X}^{n+1}_{-(i, k+2)}))\right] \nonumber \\  
\hspace{-0.2in}& \leq & (1 - p_{k+2})\left[(p_{i} - p_{k+1})C(\Pi_{n-k}(\mathbf{X}_{-(i, k+1, k+2)}^{n+1})) \right. + p_{k+1} \tilde{C}(\Pi_{n-k}(\mathbf{X}^{n+1}_{-(k+1, k+2)})) - \left. p_i \tilde{C}(\Pi_{n-k}(\mathbf{X}^{n+1}_{-(i, k+2)}))\right] \label{proof_four_b} \\ 
\hspace{-0.2in}& = & (1- p_{k+2})\left[(p_{i} - p_{k+1})C(\Pi_{n-k}(\mathbf{X}_{-(i, k+1, k+2)}^{n+1})) + p_{k+1} \tilde{C}(\Pi_{n-k}(\mathbf{X}^{n+1}_{-(k+1, k+2)})) \right. \nonumber \\
\hspace{-0.2in} & & \left. - p_i[p_{k+1} C(\Pi_{n-k-1}(\mathbf{X}^{n+1}_{-(i, k+1, k+2)})) + (1-p_{k+1}) C(\Pi_{n-k}(\mathbf{X}^{n+1}_{-(i, k+1, k+2)}))]\right] \label{proof_four_c} \\
\hspace{-0.2in}& = & (1- p_{k+2})p_{k+1}\left[\tilde{C}(\Pi_{n-k}(\mathbf{X}^{n+1}_{-(k+1, k+2)})) - p_i C(\Pi_{n-k-1}(\mathbf{X}^{n+1}_{-(i, k+1, k+2)}))\right. - \left. (1-p_i) C(\Pi_{n-k}(\mathbf{X}^{n+1}_{-(i, k+1, k+2)}))\right]  \leq 0. \label{proof_four_d} 
\end{eqnarray}

Equation (\ref{proof_four_a}) follows from the optimal ordering for computing $\Pi_{n-k}(\mathbf{X}^{n+1}_{-(i, k+1)})$, $\Pi_{n-k}(\mathbf{X}^{n+1}_{-(k+1)})$ and $\Pi_{n-k}(\mathbf{X}^{n+1}_{-i})$, which follows from the induction hypothesis for $m = n$. The inequality (\ref{proof_four_b}) follows from the induction hypothesis that $S^{(2)}_{n,k,i}(\mathbf{X}^{n+1}_{-(k+2)}) \leq 0$. Equations (\ref{proof_four_c}) and (\ref{proof_four_d}) follow from the optimal ordering for computing $\Pi_{n-k}(\mathbf{X}^{n+1}_{-(i, k+2)})$  and $\Pi_{n-k}(\mathbf{X}^{n+1}_{-(k+1, k+2)})$ respectively. $\Box$
\end{figure*}
\begin{figure*}
\subsection{Proof of Lemma \ref{lemma_one}}\label{sec_proof_one}
First, we observe that 
\begin{equation}
T_{n+1, k, i}(\mathbf{X}^{n+1}) - S^{(1)}_{n+1, k, i}(\mathbf{X}^{n+1}) = p_{k+1}\tilde{C}(\Pi_{n-k}(\mathbf{X}^{n+1}_{-(k+1)})) 
- p_{i}\tilde{C}(\Pi_{n-k}(\mathbf{X}^{n+1}_{-i})) - (p_{k+1} - p_{i})C(\Pi_{n-k}(\mathbf{X}^{n+1}_{-(k+1, i)})) \nonumber 
\end{equation}
Thus it is enough to show that 
\begin{equation}
p_{k+1}\tilde{C}(\Pi_{n-k}(\mathbf{X}^{n+1}_{-(k+1)})) - p_{i}\tilde{C}(\Pi_{n-k}(\mathbf{X}^{n+1}_{-i})) \\
\leq (p_{k+1} - p_{i})C(\Pi_{n-k}(\mathbf{X}^{n+1}_{-(k+1, i)})) \qquad \textrm{ for } i \geq k+2 \nonumber
\end{equation}
\begin{eqnarray}
& & p_{k+1}\tilde{C}(\Pi_{n-k}(\mathbf{X}^{n+1}_{-(k+1)})) - p_{i}\tilde{C}(\Pi_{n-k}(\mathbf{X}^{n+1}_{-i}))  \nonumber \\
& = &p_{k+1}\left[p_{k+2}C(\Pi_{n-k-1}(\mathbf{X}^{n+1}_{-(k+1, k+2)})) + (1-p_{k+2})C(\Pi_{n-k}(\mathbf{X}^{n+1}_{-(k+1,k+2)}))\right] \nonumber \\
& &  -p_i \left[p_{k+1}C(\Pi_{n-k-1}(\mathbf{X}^{n+1}_{-(k+1, i)})) + (1-p_{k+1})C(\Pi_{n-k}(\mathbf{X}^{n+1}_{-(k+1, i)}))\right] \label{proof_one_a} \\
& = & p_{k+1}\left[p_{k+2}C(\Pi_{n-k-1}(\mathbf{X}^{n+1}_{-(k+1,k+2)})) - p_i C(\Pi_{n-k-1}(\mathbf{X}^{n+1}_{-(k+1, i)}))\right] \nonumber \\
& & +  p_{k+1}(1 - p_{k+2})C(\Pi_{n-k}(\mathbf{X}^{n+1}_{-(k+1,k+2)})) - p_i(1-p_{k+1})C(\Pi_{n-k}(\mathbf{X}^{n+1}_{-(k+1, i)})) \nonumber \\
&  \leq & p_{k+1}\left[ (1-p_i)C(\Pi_{n-k}(\mathbf{X}^{n+1}_{-(k+1, i)})) - (1-p_{k+2})C(\Pi_{n-k}(\mathbf{X}^{n+1}_{-(k+1,k+2)}))\right] \nonumber \\
& & + p_{k+1}(1 - p_{k+2})C(\Pi_{n-k}(\mathbf{X}^{n+1}_{-(k+1,k+2)})) - p_i(1-p_{k+1})C(\Pi_{n-k}(\mathbf{X}^{n+1}_{-(k+1, i)})) \hspace{0.2in}\label{proof_one_b} \\
&=& (p_{k+1} - p_i)C(\Pi_{n-k}(\mathbf{X}^{n+1}_{-(k+1, i)})) \nonumber
\end{eqnarray}
Equation \ref{proof_one_a} follows from the optimal order for computing $\Pi_{n-k}(\mathbf{X}^{n+1}_{-(k+1)})$ and $\Pi_{n-k}(\mathbf{X}^{n+1}_{-i})$. The inequality in \ref{proof_one_b} follows from the induction hypothesis $T_{n, k, i}(\mathbf{X}^{n+1}_{-(k+1)}) \leq 0$. $\Box$
\end{figure*}
\begin{figure*}
\subsection{Proof of Lemma \ref{lemma_two}}\label{sec_proof_two}
First, we observe that 
\begin{equation}
T_{n+1, k, i}(\mathbf{X}^{n+1}) - S^{(2)}_{n+1, k, i}(\mathbf{X}^{n+1}) = (1-p_{k+1})\tilde{C}(\Pi_{n-k+1}(\mathbf{X}^{n+1}_{-(k+1)})) 
- (1-p_{i})\tilde{C}(\Pi_{n-k+1}(\mathbf{X}^{n+1}_{-i})) - (p_{i} - p_{k+1})C(\Pi_{n-k}(\mathbf{X}^{n+1}_{-(i,k+1)})) \nonumber 
\end{equation}
Thus it is enough to show that 
\begin{equation}
(1-p_{k+1})\tilde{C}(\Pi_{n-k+1}(\mathbf{X}^{n+1}_{-(k+1)})) - (1-p_{i})\tilde{C}(\Pi_{n-k+1}(\mathbf{X}^{n+1}_{-i})) 
\leq (p_{i} - p_{k+1})C(\Pi_{n-k}(\mathbf{X}^{n+1}_{-(i,k+1)})) \qquad \textrm{ for } i \leq k \nonumber
\end{equation}
\begin{eqnarray}
& &  (1-p_{k+1})\tilde{C}(\Pi_{n-k+1}(\mathbf{X}^{n+1}_{-(k+1)})) - (1-p_{i})\tilde{C}(\Pi_{n-k+1}(\mathbf{X}^{n+1}_{-i})) \nonumber \\
&  = &  (1-p_{k+1})\left[p_{k}C(\Pi_{n-k}(\mathbf{X}^{n+1}_{-(k, k+1)})) + (1-p_{k})C(\Pi_{n-k+1}(\mathbf{X}^{n+1}_{-(k,k+1)}))\right] \nonumber \\
& &  -(1- p_i) \left[p_{k+1}C(\Pi_{n-k}(\mathbf{X}^{n+1}_{-(i, k+1)})) + (1-p_{k+1})C(\Pi_{n-k+1}(\mathbf{X}^{n+1}_{-(i,k+1)}))\right] \label{proof_two_a} \\
& = & (1- p_{k+1})\left[(1- p_{k})C(\Pi_{n-k+1}(\mathbf{X}^{n+1}_{-(k,k+1)})) - (1-p_i) C(\Pi_{n-k+1}(\mathbf{X}^{n+1}_{-(i, k+1)}))\right] \nonumber \\
& &  + p_{k}(1 - p_{k+1})C(\Pi_{n-k}(\mathbf{X}^{n+1}_{-(k,k+1)})) - p_{k+1}(1-p_{i})C(\Pi_{n-k}(\mathbf{X}^{n+1}_{-(i,k+1)})) \nonumber \\
& \leq & (1-p_{k+1})\left[ p_i C(\Pi_{n-k}(\mathbf{X}^{n+1}_{-(i,k+1)})) - p_k C(\Pi_{n-k}(\mathbf{X}^{n+1}_{-(k,k+1)}))\right] \nonumber \\
& & + p_{k}(1 - p_{k+1})C(\Pi_{n-k}(\mathbf{X}^{n+1}_{-(k,k+1)})) - p_{k+1}(1-p_{i})C(\Pi_{n-k}(\mathbf{X}^{n+1}_{-(i,k+1)})) \label{proof_two_b} \\
&  =  & (p_{i} - p_{k+1})C(\Pi_{n-k}(\mathbf{X}^{n+1}_{-(i,k+1)})) \nonumber
\end{eqnarray}
Equation \ref{proof_two_a} follows from the optimal order for computing $\Pi_{n-k+1}(\mathbf{X}^{n+1}_{-(k+1)})$ and $\Pi_{n-k+1}(\mathbf{X}^{n+1}_{-i})$. The inequality in \ref{proof_two_b} follows from the induction hypothesis $T_{n, k-1, i}(\mathbf{X}^{n+1}_{-(k+1)}) \leq 0$. $\Box$.
\end{figure*}
\bibliographystyle{unsrt}
\bibliography{isit_biblio}
\end{document}